\documentclass[12pt,twoside,leqno]{amsart}
\usepackage{mathptmx}
\usepackage{amssymb, amsmath, amsthm, enumerate,xfrac}
\usepackage{tabularx,floatrow}
\usepackage{graphicx}
\makeatletter
\renewcommand\section{\@startsection {section}{1}{\z@}%
                                   {-3.5ex \@plus -1ex \@minus -.2ex}%
                                   {2.3ex \@plus.2ex}%
                                   {\normalfont\large\bfseries}}
\renewcommand\subsection{\@startsection{subsection}{2}{\z@}%
                                     {-3.25ex\@plus -1ex \@minus -.2ex}%
                                     {1.5ex \@plus .2ex}%
                                     {\normalfont\large\bfseries}}
\renewcommand\subsubsection{\@startsection{subsubsection}{3}{\z@}%
                                     {-3.25ex\@plus -1ex \@minus -.2ex}%
                                     {1.5ex \@plus .2ex}
																		{\normalfont\large\bfseries}}%
\renewcommand{\@secnumfont}{\bfseries}
\makeatother
\DeclareFloatFont{small}{\footnotesize}
\begin{document}
\setcounter{page}{1}
\vspace*{2.0cm}

\title[Closed-form bond price] {A closed-form formula for pricing bonds between coupon payments}
\author[Sylvia Gottschalk] {Sylvia Gottschalk$^{1}$}
\date{}
\maketitle

\vspace*{-0.2cm}

\begin{center}
{\footnotesize  $^1$Middlesex University, London NW4 4BT, United Kingdom\\
}
\end{center}
\vskip 2mm

\noindent{\renewcommand\baselinestretch{1.0}\selectfont
{\tiny  Copyright \copyright\ 2018  Sylvia Gottschalk. This is an open access article distributed under the Creative Commons Attribution License, which permits unrestricted use, distribution, and reproduction in any medium, provided the original work is properly cited. } }\par\vskip 2mm

{\footnotesize 

\noindent {\bf Abstract.}  We derive a closed-form formula for computing bond prices between coupon payments. Our results cover both the `Treasury' and the `Street' pricing methods used by sovereign and corporate issuers. We apply our formulas to two UK gilts, the 8\% Treasury Gilt 2015, and the 0\sfrac{1}{2}\% Treasury Gilt 2022, and show that we can obtain the dirty price of these bonds at any date with a minimum of calculations, and without intensive computational resources. \vskip 1mm

\vskip 1mm

\emph{This paper has been accepted for publication in Mathematical Finance Letters.}

}

\vskip 6mm

\renewcommand{\thefootnote}{}
\footnotetext{ $^*$ Corresponding author   
\par
E-mail address: s.gottschalk@mdx.ac.uk
\par
Received April 11, 2018

}

\section{Introduction}

Bond pricing is a basic feature of fixed-income analytics, and is a direct application of the concept of time value of money. In the existing literature, most fixed-income securities are priced at the issuance date, $t=0$ by convention. However, the current framework cannot be directly applied to pricing bonds traded after they were issued unless the date bonds exchange ownership coincide with a coupon payment date. When bonds are traded between coupon payments, the conventional formulas cannot be applied, and fixed-income analysts usually rely on `back-of-the-envelope' calculations of \emph{dirty} and \emph{clean} prices.\\   

Advanced fixed income analytics do address the issue, albeit inconsistently. [6],[7],[8],[3], and even the very advanced [9] briefly cover the topic and present pricing formulas that are either complex or cumbersome. Further, the treatment of the calculations of dirty price and clean price is frequently counter-intuitive, since they are based on simple interest.\\

This paper presents a simple closed-form formula for bond pricing between coupon payments that derives from first principles and is theoretically correct. Our results are more general than the current framework, and we prove that we can retrieve the conventional formula for pricing  bonds at coupon dates as a special case. We also demonstrate that bond traders' `dirty price' effectively assumes that interest between coupon payment is simple interest, when mathematical consistency requires that all interest should be coumpounded.\\ 

We illustrate our results with an application to two UK government bonds, the 8\% Treasury Gilt 2015, and the 0\sfrac{1}{2}\% Treasury Gilt 2022. We show that the implmentation of our results involves very few steps irrespective of the maturity of the bonds. Existing approaches such as [1] and [8] become difficult to use for pricing bonds with a maturity longer than 2 years (semi-annual coupon payments) or four years (annual payments). Although absolutely correct, at longer maturities, they require the laborious calculation of several discount ratios and become rapidly computationally inefficient.\\ 

The main results of the paper will be derived in Section \ref{sec:back}. A detailed application of our results to the UK 8\% Treasury Gilt 2015 and the  0\sfrac{1}{2}\% Treasury Gilt 2022 in Section \ref{sec:illus} shows that our formulas can replicate actual bond market practice. Section \ref{sec:conc} concludes.

\section{The price of bonds between interest payments \label{sec:back} }

The fair price of a bond is the sum of the present value of the cash flow of coupon payments and the present value of the principal. The price is usually calculated at the issuance date (time $t=0$ by convention). If a bond is bought at $t=0$ and hold onto until maturity, $t=N$, the buyer receives all the coupon payments between $t=0$ and $t=N$. However, bonds may be traded at any time before maturity, and should the transaction date fall between two coupon payments, the new buyer will receive the full interest payment at the next coupon date.\\

\begin{figure}
\begin{center}
\resizebox*{14cm}{!}{\includegraphics{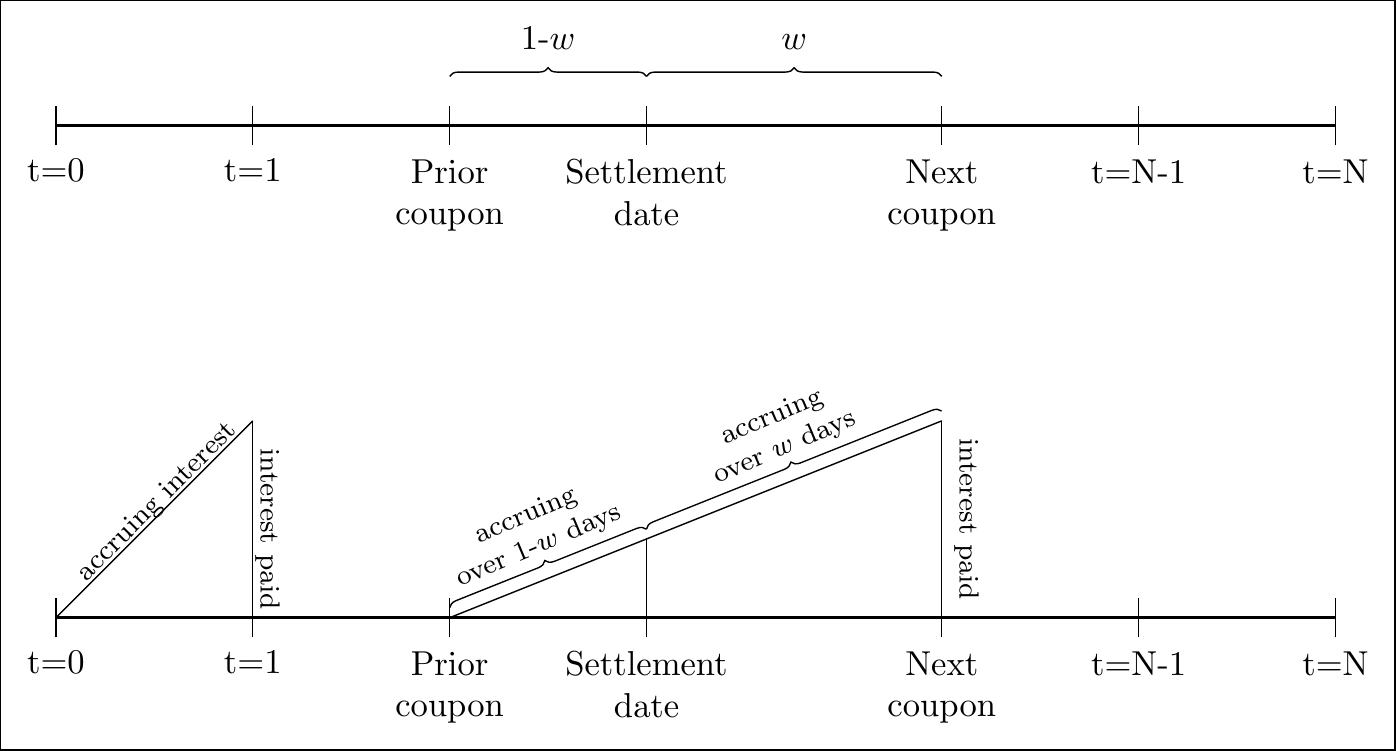}}
\end{center}\caption{Accrued interest of a coupon bond\label{f1}}
\end{figure}

Figure \ref{f1} illustrates the issue. The upper graph shows a coupon period split into two fractions by the settlement date, $w$, and $1-w$, with $0<w\leq 1$. $w$ is equal to the ratio of the number of days between the settlement date and the next coupon date to the total number of days in the coupon period (see Section \ref{sub:accrued} for details). The new owner receives the totality of the interest payment at time $t$=\emph{Next coupon}, but is only entitled to the interest compounded over $w$-th of the coupon period. The lower graph shows that interest accrues daily until the coupon payment date, e.g., at times $t=1$ and $t$=\emph{Next coupon}, when it is then paid in full.\\  

Two questions arise. The first concerns the fair price of a bond bought within the coupon period. The second relates to the compensation due to the seller of the bond for the loss of interest payment in the period between the previous coupon and the settlement date. The next two subsections address these questions.

\subsection{Bond pricing\label{sec:main}}

This section shows how to calculate the fair price of a coupon bond at time $t=w$. For the sake of simplicity, we will derive the main results assuming coupons are paid annually. The formulas for other frequencies are presented when needed.\\

Intuitively, one would want to take the extended formula for calculating the fair price at time $t=0$ with first coupon payment at $t=1$,

\begin{equation}
P=\sum_{t=1}^{N}\frac{C}{(1+y)^{t}}+\frac{M}{(1+y)^{N}}\label{old}
\end{equation}

where,\\
$C$= coupon payment value\\
$N$= number of years\\
$y$= yield to maturity or discount rate\\
$M$= face value of the bond at maturity\\

and change the starting date to $t=w$, i.e., 

\begin{equation}
P=\sum_{t=w}^{N}\frac{C}{(1+y)^{t}}+\frac{M}{(1+y)^{N}}\label{old1}
\end{equation}

Although (\ref{old1}) captures the spirit of the formula we wish to obtain, it suffers from three shortcomings. First, the actual value of $w$ depends on the problem at hand, i.e., $w$ can be 1/5-th, or 0.7-th of the coupon period, or any value between 0 and 1. As a result, in (\ref{old1}), unlike in (\ref{old}), the starting point of the sum of discounted cash flows is not known. Secondly, and relatedly, the counter of the sum is an integer by convention to make clear the sequence of powers in the discount factor. Specifically, if a coupon payment is made at time $t=10$, and $t$ can only be a whole number, the next coupon payment is necessarily at $t=11$. If $t=w=0.8667$, the date of the next payment is not immediately obvious. Finally, (\ref{old}) and (\ref{old1}) are not equivalent. In (\ref{old}), $t=1$ is the date of the first coupon payment, whilst in (\ref{old1}) $t=w$ is the date interest starts to accrue, as can be seen in the lower graph of Figure \ref{f1}. This last point can be tackled by two distinct methods, which will be detailed below.

\subsubsection{`Street' method\label{sub:street}}  

The `Street' method is commonly used by corporate issuers (see [1],[3],[7],[8]). 

\noindent {\bf Theorem 2.1.} {\it Let $y,M $ and $i \in (0;\infty)$, $w \in (0,1]$, and let $N\in \mathbb{N}$, then the fair price of a coupon bond between coupon payments is  

\begin{equation}
P=\frac{1}{(1+y)^{w}}\left\{M\,i\left[\frac{1-\frac{1}{(1+y)^{N}}}{1-\frac{1}{(1+y)}}\right]
+\frac{M}{(1+y)^{N-1}}\right\}\label{final}
\end{equation}
}

\vskip 2mm
\begin{proof} Let $q\equiv \frac{1}{(1+y)}$, then

\begin{equation}
P=q^{w}\left\{M\,i\sum_{t=1}^{N} q^{t-1}+M\, q^{N-1}\right\}\label{basic1}
\end{equation}

The sum can be rewritten as

\begin{equation}
S=\sum_{t=1}^{N} q^{t-1}\label{sum1}
\end{equation}

and can be expanded to

\begin{equation}
S=q^{0}+q^{1}+q^{2}+...+q^{N-1}\label{sum1a}
\end{equation}

Multiplying both sides of (\ref{sum1a}) by $q$,

\begin{equation}
qS=q^{1}+q^{2}+...+q^{N-1}+q^{N}\label{sum1b}
\end{equation}

and subtracting (\ref{sum1b}) from (\ref{sum1a}), we find

\begin{equation}
S=\frac{1-q^{N}}{1-q}\label{sum1d}
\end{equation}

Substituting (\ref{sum1d}) into (\ref{basic1}) gives the closed form version of the bond price formula,

\begin{equation}
P=q^{w}\left\{M\,i\,\frac{1-q^{N}}{1-q}+M\,q^{N-1}\right\}\label{basic1a}
\end{equation}

Finally, by replacing $q$=$1/(1+y)$ into (\ref{basic1}) yields (\ref{final}).
\end{proof}

\noindent {\bf Corollary 2.1.1} {\it  If $w=1$, (\ref{final}) reverts to

\begin{equation}
P=M\,i\left[\frac{1-\frac{1}{(1+y)^{N}}}{y} \right]+\frac{M}{(1+y)^{N}}\label{finala1}
\end{equation} 
}

\vskip 2mm

\begin{proof}

\begin{equation}
P=q^{1}\left\{M\,i\left[\frac{1-q^{N}}{1-q}\right]+M\,q^{N-1}\right\}\label{basica1a}
\end{equation}

and multiplying through $q$, we obtain

\begin{equation}
P=q\,M\,i\left[\frac{1-q^{N}}{1-q}\right]+M\,q^{N}\label{basica1b}
\end{equation}

factorising $1-q$ as $q\,(1/q-1)$ and replacing this into (\ref{basica1b}) gives,
 
\begin{equation}
P=q\,M\,i\left[\frac{1-q^{N}}{q\,(1/q-1)}\right]+M\,q^{N}\label{basica1c}
\end{equation}

cancelling out $q$, and replacing  $q\equiv \frac{1}{(1+y)}$,

\begin{equation}
P=M\,i\left[\frac{1-\frac{1}{(1+y))^{N}}}{1+y-1}\right]+\frac{M}{(1+y)^{N}}
\label{basica1c}
\end{equation}

(\ref{finala1}) follows trivially.
\end{proof}

(\ref{finala1}) is the price of a coupon bond bought at a coupon payment date. It is the closed-form of (\ref{old}).\\

\subsubsection{`Treasury' method\label{sub:t}}

Sovereign issuers, such as the  UK Debt Management Office (DMO), tend to favour the procedure established by the International Securities Market Association (ISMA), (see [2] and [4]). [7] and [3] refer to this variant as the `Treasury' method. It is based on the assumption that the present value of cash flows should start at the time interest accrues rather than at the time of the first coupon payment.

\noindent {\bf Theorem 2.1.2} {\it Let $y, M$ and $i \in (0;\infty)$, $w \in (0,1]$, and let $N\in \mathbb{N}$, then the fair price of a coupon bond between coupon payments under the `Treasury' method is

\begin{equation}
P=q^{w}\left\{M\,i\,\frac{1-q^{N+1}}{1-q}+M\,q^{N}\right\}\label{tbasic}
\end{equation}
where $q=\frac{1}{1+y}$
}

\begin{proof}
Let $q\equiv \frac{1}{1+y}$, then,

\begin{equation}
P=q^{w}\left\{M\,i\sum_{t=0}^{N} q^{t}+M\, q^{N}\right\}\label{treasury1}
\end{equation} 

The sum can be expanded to

\begin{equation}
S=q^{0}+q^{1}+q^{2}+...+q^{N}\label{tsum1a}
\end{equation}

Multiplying both sides of (\ref{tsum1a}) by $q$,

\begin{equation}
qS=q^{1}+q^{2}+...+q^{N}+q^{N+1}\label{tsum1b}
\end{equation}

and subtracting (\ref{tsum1b}) from (\ref{tsum1a}), we find

\begin{equation}
S=\frac{1-q^{N+1}}{1-q}\label{tsum1d}
\end{equation}

Substituting (\ref{tsum1d}) into (\ref{treasury1}) yields (\ref{tbasic}).
\end{proof}

The UK Debt Management Office has developed a variant of (\ref{tbasic}),  

\begin{equation}
P=q^{w}\left\{C_1+C_2\,q+C\frac{q^2(1-q^{N-1})}{(1-q)}+M\,q^{N}\right\}\label{isma}
\end{equation}

A cursory comparison of (\ref{tbasic}) and (\ref{isma}) shows that these two formulas are related. In fact, (\ref{isma}) is also obtained from (\ref{treasury1}), by `extracting' $C_1+C_2\,q$ from the sum, which then starts at time $t=2$. Replacing this sum by its closed-form results in (\ref{isma}). Nonetheless, the difference is not merely cosmetic. It has a financial justification that will be detailed in the next section, where we illustrate the application of the formulas above to two UK government bonds, the 8\% Treasury Gilt 2015, and the  0\sfrac{1}{2}\% Treasury Gilt 2022. \\

\subsection{Accrued interest \label{sub:accrued}}

Having found the fair price of a bond, we now turn to evaluating the amount of accrued interest bond sellers should receive from the buyers to compensate for loss of interest. Figure \ref{f1} shows that it should be the amount of interest accrued during $1-w$ days, and internal consistency requires that it should be compounded. We present below the universally accepted market practice and prove that this practice is based on simple interest.\\

Assuming coupons are paid annually, the number of days in the coupon period is 365 under the `Actual/Actual' day count convention\footnote{There are many day count conventions in Finance. The most common are `Actual/Actual', `30/360', and `Actual/360'. The first is the calendar month/year. The second postulates that any month of the year has 30 days, and the year 360. The third assumes months have the number of days in the calendar, but the year only 360 days. The choice of day count convention is not innocuous. More interest is paid under the `Actual/Actual' than under the `30/360'.}. If, for instance, the settlement date in Figure \ref{f1} is the 1rst April 2018, and the next coupon date 31 December 2018, there are 274 days in this period of time. 274 days represent a fraction $w=274/365$ of the calendar year 2018. The period of time between the previous coupon and the settlement date has 90 days, and accounts for a fraction $1-w=91/365=(365-91)/365$ of 2018. The accrued interest to be paid to the bond seller is given by      

\begin{equation}
\textrm{Accrued interest}=\textrm{coupon value}\times\frac{\textrm{number of days since last coupon}}
{\textrm{number of days in coupon period}}\label{ai}
\end{equation}

The coupon value is given by multiplying the face value of the bond by the coupon rate. Let $i$ be the annual coupon rate and $M$ the face value of the bond. Substituting in (\ref{ai}), the accrued interest becomes

\begin{equation}
\textrm{AI}=M\times i\times\frac{\textrm{number of days since last coupon}}
{\textrm{number of days in coupon period}}\label{aia1}
\end{equation}

The ratio is merely a fraction of a year, i.e., it represents time. (\ref{aia1}) can be read as 

\[\textrm{AI}=M\times i\times t\]

where $t=1-w$. This is the textbook definition of simple interest. The fixed-income industry version of the bond prices between coupon payments we derived in section \ref{sec:main} is 

\begin{equation}
\textrm{Dirty Price}=\textrm{Clean Price}+\textrm{Accrued Interest}\label{dirty}
\end{equation}

where the accrued interest is given by (\ref{aia1}). The clean price is normally the market quote at the end of the business day. It should be emphasized that all the formulas derived in section \ref{sec:main} \emph{are} dirty prices. However, the results produced by (\ref{basic1a}), (\ref{tbasic}) and (\ref{isma}) and those produced by (\ref{dirty}) will invariably differ. (\ref{dirty}) suffers from the fundamental flaw of being based on simple interest, whereas in (\ref{basic1a}), (\ref{tbasic}) and (\ref{isma}) bond coupon interest is always calculated on a compounded basis. 

\section{Illustration: UK 8\% Treasury Gilt 2015 and  0\sfrac{1}{2}\% Treasury Gilt 2022\label{sec:illus}}

The auction press notice for the UK 8\% Treasury Gilt 2015\footnote{The press notices for these bonds are the files prosp160796a.pdf and pr110417.pdf, respectively. Both files can be downloaded from the DMO website. In this section we will use `interest' and `coupon' interchangeably.} informs that this bond pays 8\% interest semi-annually on 7 June and 7 December. The maturity date is 25 July 2015 and the issue date 25 July 1996. The 0\sfrac{1}{2}\% Treasury Gilt 2022 is a 0.5\% coupon bond with maturity date 22 July 2022, issued on 21 April 2017. The coupon is payable semi-annually on 22 January and 22 July of each year until maturity. The face value of both bonds is \pounds 100.\\

\subsection{The UK 8\% Treasury Gilt 2015\label{sub:gilt2015}}

Table \ref{tab:gilt2015} presents four scenarios that differ by their settlement dates, 24-May-99, 26-May-99, 27-May-99 and 07-June-99. The variables $C_1$, $C_2$, and $C$ are the semi-annual coupon payments, calculated as $100\times (8\%/2)=4$. $N$ is the number of coupon payments, $r$ is the number of days between the settlement date and the date of the next coupon payment, whilst $s$ is the number of days in the coupon period. Our variable $w$ is equal to the ratio $r/s$. The day count convention used by the DMO is `Actual/Actual', signifying that the number of days in a month depends on the calendar month and that the number of days in the year is 365 (see [4]). $q=\frac{1}{1+yield/2}=\frac{1}{1+0.02225}=0.978258211$. \\
 
\begin{table}
  \caption{Dirty price of the 8\% Treasury Gilt 2015 (in pounds)}
\begin{center}
    \begin{tabular}{lcccc}    \hline
  Scenario&1&2&3&4\\
	Bond&8\% 2015&8\% 2015&8\% 2015&8\% 2015\\
	Yield &0.04445&0.04445&0.04445&0.04445\\
	Settlement date&24-May-99&26-May-99&27-May-99&07-Jun-99\\
	Ex-dividend date&26-May-99&26-May-99&26-May-99&26-May-99\\
	Ex-dividend&No&No&Yes&No\\
	Previous quasi-coupon date&07-Dec-98&07-Dec-98&07-Dec-98&07-Dec-98\\
	Next quasi-coupon date&07-Jun-99&07-Jun-99&07-Jun-99&07-Dec-99\\
	$C_1$&4&4&4&4\\
	$C_2$&4&4&4&4\\
	C&4&4&4&4\\
	q&0.978258211&0.978258211&0.978258211&0.978258211\\
	r&14&12&11&183\\
	s&182&182&182&183\\
	Maturity date&07-Dec-15&07-Dec-15&07-Dec-15&07-Dec-15\\
	N&33&33&33&33\\
	Dirty price&145.012268&145.047301&141.070132&141.257676\\ \hline
	\end{tabular}%
\end{center}
\scriptsize{Source: Debt Management Office (DMO), https://www.dmo.gov.uk/media/15009/yldconv.pdf. The notation of some variables was changed to conform to the notation of our paper.}
  \label{tab:gilt2015}%
\end{table}%

The only variables that are specific to this table - and to the Debt Management Office - are the dividend and the quasi-coupon date. The dividend is simply the interest payment, whilst `ex-dividend' refers to the week preceding the interest payment date. The DMO pays interest to the registered holder of the bond during these seven days. If the bond is sold during the ex-dividend period, the seller will receive the full amount of interest, but will have to refund some back to the buyer of the bond (see [5] pp.15-16) Finally, the `quasi-coupon' date is the day compounding occurs, irrespective of whether a payment is made (see [4]). Given that coupon payments are made over seven days, it seems sensible to distinguish between the payment date and the compounding date. However, this distinction is particularly relevant for a gilt issued between the dates that will constitute its coupon period. For instance, the 0\sfrac{1}{2}\% Treasury Gilt 2022 was issued on 21 April 2017, but pays interest on 22 January and 22 July. As will be seen below, in order to calculate the accrued interest on any day before 22 July 2017, the DMO assumes that interest compounding started on 22 January 2017, i.e., before the gilt was actually issued. The 22nd January 2017 is thus a `quasi-coupon' date.\\[3mm]

\underline{Scenario 1}\\

Given that $w=r/s=0.07692308$, $q^w=0.9983105345$, $q^{N}=0.4841339743$, $q^{N-1}=0.494839386$, and $q^2=0.9569891274$. Substituting these values in formula (\ref{isma}) gives

\[
P=0.9983105345 \Bigl(4+4\times 0.978258211+4\frac{(0.9569891274)(0.505160614)}{(0.021741789)}+\\
\]
\[100\times 0.4841339743 \Bigr)
\]

from which,

\[
P=0.9983105345\left(7.913032844+88.94083468+48.41339743\right)=145.02184\label{sce1a}
\]

This is the dirty price for Scenario 1 seen in Table \ref{tab:gilt2015}. Scenario 2 is merely a repetition of the above, with the minor difference that the settlement date coincides with the ex-dividend date. The values to be inserted in (\ref{isma}) will be different but, unlike Scenario 3, there are no modifications of the formula.\\

\underline{Scenario 3}\\

Under this scenario, the settlement date is within the ex-dividend period. As explained above, the seller has already been paid the coupon interest and has to refund it to the buyer.  Consequently, the interest payment $C_1$ is excluded from (\ref{isma}), and the price of the gilt is 

\[
P=0.998672322 \Bigl(4\times 0.978258211+4\frac{(0.9569891274)(0.505160614)}{(0.021741789)}+\\ \]
\[
100\times 0.4841339743 \Bigr)
\]

where $w=11/182=0.06043956$ and $q^w=0.998672322$. 

\[
P=0.998672322\left(3.913032844+88.93124549+48.41339743\right)=141.070132
\]      

The formula we derived in \ref{sub:t} can directly replicate the dirty prices published by the DMO in Table \ref{tab:gilt2015}, except in cases where the coupon payment is made before the coupon date. Clearly, substituting in (\ref{tbasic}) the values for $q$, $w$, $C_1$, $C_2$, $C$, and $N$ from   Scenario 1, we obtain
 
\[
P=0.9983105345(96.84427797+48.41339743)=145.012268
\]

Although our formula (\ref{tbasic}) does not give the results of Scenario 3 directly, it is very easy to retrieve the DMO price by simply subtracting from (\ref{tbasic}) the value of the coupon payment multiplied by $q^w$,

\[
P=0.998672322 \Bigl(96.84427797+48.41339743\Bigr)-4\times 0.998672322=141.07013\] 

\subsection{0\sfrac{1}{2}\% Treasury Gilt 2022\label{sub:gilt2022}}

\begin{table}
\begin{center}
  \caption{Dirty price, Clean price and accrued interest of the 0\sfrac{1}{2}\% Treasury Gilt 2022 (in pounds)}
    \begin{tabular}{llllrl}    \hline
    Gilt Name &  Date & Clean Price& Dirty Price & Accrued Interest& Yield (\%) \\ \hline
    0\sfrac{1}{2}\% Treasury Gilt 2022 & 03-Jul-17 & 99.04 & 99.26514 & 0.225138 & 0.693781 \\
    0\sfrac{1}{2}\% Treasury Gilt 2022 & 04-Jul-17 & 99.18 & 99.40652 & 0.226519 & 0.665481\\
    0\sfrac{1}{2}\% Treasury Gilt 2022 & 05-Jul-17 & 99.28 & 99.5079 & 0.227901 & 0.645297 \\
    0\sfrac{1}{2}\% Treasury Gilt 2022 & 06-Jul-17 & 98.94 & 99.16928 & 0.229282 & 0.714435\\
    0\sfrac{1}{2}\% Treasury Gilt 2022 & 07-Jul-17 & 99.08 & 99.31343 & 0.233425 & 0.686271\\
    0\sfrac{1}{2}\% Treasury Gilt 2022 & 10-Jul-17 & 99.24 & 99.47481 & 0.234807 & 0.653822 \\
    0\sfrac{1}{2}\% Treasury Gilt 2022 & 11-Jul-17& 99.16 & 99.39619 & 0.236188 & 0.670183 \\
    0\sfrac{1}{2}\% Treasury Gilt 2022 & 12-Jul-17 & 99.19 & 99.42757 & 0.237569 & 0.664167\\
        \hline
    \end{tabular}%
		\end{center}
\scriptsize{Source: Debt Management Office (DMO). http://www.dmo.gov.uk/data/ExportReport?reportCode=D3B.}
  \label{tab:gilt2022}%
\end{table}%

The results of the second case study are presented in Table \ref{tab:gilt2022a}, which summarizes the application of formulas (\ref{basic1a}), (\ref{tbasic}) and (\ref{isma}) to calculate the dirty price of the 0\sfrac{1}{2}\% gilt between 03/07/2017 and 12/07/2017. (\ref{ai}) was used to obtain the values in the column `accrued interest'. The last two columns show the difference between the published dirty price seen in Table \ref{tab:gilt2022} and the calculated dirty prices in Table \ref{tab:gilt2022a}. The number of days between the settlement date and the next coupon date (22 July 2017) is in column `$r$'. `$w$ and `$q$ are defined as in section \ref{sub:gilt2015}. The number of days in the coupon period, $s$, is 181, and is the number of days between 22 January 2017 and 22 July 2017. The coupon payment is $100\times (0.5\%/2)=0.25$. The accrued interest on 03/07/2017 is\footnote{the numerator is $1-w$, and 1 day must be added because the 03/07/2017 is included in the accrued interest.}

\[0.25\times \frac{181-19+1}{181}=0.25\times \frac{163}{181}=0.25\times 0.9005524862=0.225138\]

which is precisely the accrued interest published by the DMO. This also confirms that the `quasi-coupon' date 22 January 2017 is indeed the starting date of the coupon period, even though it precedes the issue date of the gilt, 21 April 2017. If 181 is replaced by the number of days between 21 April 2017 and 03/07/2017, 91, the resulting accrued interest would be \pounds 0.2005. The dirty price seen in Table \ref{tab:gilt2022} is found by adding the accrued interest to the clean price, namely, $99.04+0.225138=99.265138$. \\

The dirty prices found using our closed-form (\ref{tbasic}) and (\ref{isma}) are in columns 6 and 7. They are unsurprisingly identical. We have seen while analysing the UK 8\% Treasury Gilt 2015 that in the `ex-dividend' period the DMO formula and ours produce exactly the same values. The range of data chosen for the 0\sfrac{1}{2}\% Treasury Gilt 2022 precedes the `ex-dividend' period, which would start on 17/07/2017.\\

Finally, the comparison between the calculated dirty prices and those published by the DMO shows a difference of about 20\% between the two `Treasury methods' and actual data, whilst the difference between the `Street method' and the data is less than 10\%. This is very interesting and means that the DMO's theoretical dirty price formula cannot replicate its own data. However, this discrepancy can be accounted for by two factors. First, the published dirty prices are calculated using the market practice delineated in Section \ref{sub:accrued}. So, it should not be surprising that the `Street method' replicated more closely data created using the market practice. Second, the dirty price seen in Table \ref{tab:gilt2022} depends on the value of the clean price, via equation (\ref{dirty}). The clean price is merely the market quote at the end of the business day, and depends on supply and demand for bonds on the day, and on other market conditions. Consequently, any price obtained from (\ref{dirty}) will reflect those market conditions. Nonetheless, the two case studies presented in this section confirm that the `Treasury' closed-form we derived is equivalent to that of the Debt Management Office. 

\begin{table}[htbp]
\caption{Calculated dirty price and accrued interest of the 0\sfrac{1}{2}\% Treasury Gilt 2022 (in pounds) }
    \begin{tabular}{llccp{1.3cm}p{1.3cm}p{1.3cm}p{1.3cm}p{1.5cm}p{1.4cm}}    \hline
   Date  & r     & w     & q     & Dirty Price (\ref{basic1a}) & Dirty price (\ref{tbasic}) & Dirty Price (\ref{isma}) & Accrued Interest (\ref{ai}) & Difference DMO-(\ref{isma}) & Difference DMO-(\ref{basic1a}) \\ \hline
    03/07/2017 & 19    & 0.104972 & 0.996543 & 99.17000 & 99.077089 & 99.077089 & 0.225138 & 0.188049 & 0.095136 \\
    04/07/2017 & 18    & 0.099448 & 0.996684 & 99.32495 & 99.245467 & 99.245467 & 0.226519 & 0.161052 & 0.081571 \\
    05/07/2017 & 17    & 0.093923 & 0.996784 & 99.43603 & 99.366161 & 99.366161 & 0.227901 & 0.141740 & 0.071867 \\
    06/07/2017 & 16    & 0.088398 & 0.996441 & 99.06427 & 98.961574 & 98.961574 & 0.229282 & 0.207708 & 0.105014 \\
    07/07/2017 & 15    & 0.082873 & 0.996580 & 99.21812 & 99.128763 & 99.128763 & 0.230663 & 0.184662 & 0.095303 \\
    10/07/2017 & 12    & 0.066298 & 0.996742 & 99.39883 & 99.324889 & 99.324889 & 0.234807 & 0.149918 & 0.075977 \\
    11/07/2017 & 11    & 0.060773 & 0.996660 & 99.31235 & 99.230623 & 99.230623 & 0.236188 & 0.165565 & 0.083839 \\
    12/07/2017 & 10    & 0.055249 & 0.996690 & 99.34662 & 99.267750 & 99.267750 & 0.237569 & 0.159819 & 0.080951 \\
    \hline
    \end{tabular}
 \label{tab:gilt2022a}%
\end{table}%

\section{Conclusion\label{sec:conc}}

This paper has presented an innovative and theoretically coherent way to price fixed-income securities between coupon payments. Our results cover the two main exisitng pricing methods, `Street' and `Treasury'. We proved that the industry's \emph{dirty price} calculated by market fixed-income analysts, and commonly found on platforms such Bloomberg, is theoretically inconsistent. Our results were derived from first principles and replicate the more rigorous framework of the UK Debt Management Office.

\noindent{\bf Conflict of Interests}

\noindent The authors  declare  that there is no conflict of interests.


\vskip 2mm

\begin{thebibliography}{99}
\bibitem{1} G.V. Boyles, T.W. Secrest and R.B. Burney, The pricing of bond between coupon payments: From theory to practice,
Journal of Economics and Finance Education 2(4) (2005), 61-68.
\bibitem{2}  P.J. Brown, Bond markets: Structure and yield calculations, ISMA and Glamour Drummond Publishing (GDP), (1998), Cambridge, United Kingdom.
\bibitem{3} M. Choudhry, The bond and money markets: {S}trategy, trading, analysis, ButterWiley-Heinemann, Bath, United Kingdom, (2003).
\bibitem{4} Debt Management Office (DMO), Formulae for Calculating Gilt Prices from Yields, 3rd Ed., (2005), https://www.dmo.gov.uk/media/1955/yldeqns.pdf
\bibitem{5} Debt Management Office (DMO), A private investor's guide to gilts, 4th Ed., (2004), http://www.dmo.gov.uk/media/14702/pig201204.pdf
\bibitem{6} F.J. Fabozzi, Bond markets, analysis, and strategies, 8th ed., Pearson, (2013)
\bibitem{7} F.J. Fabozzi and M. Choudhry, The handbook of European fixed-income securities, Wiley, New Jersey, (2004).
\bibitem{8} F.J. Fabozzi and S. V. Mann, Introduction to fixed income analytics,FJF Frank J. Fabozzi Associates, New Hope, Pennsylvania,  (2001).
\bibitem{9} P. Veronesi, Fixed income securities: {V}aluation, risk, and risk management, Wiley (2010). 
\end{thebibliography}
\end{document}